\documentclass{egpubl}
\usepackage{egsgp2021}
 
\SpecialIssuePaper         

\usepackage[T1]{fontenc}
\usepackage{dfadobe}  
\usepackage{float} 

\usepackage{cite}  
\BibtexOrBiblatex
\electronicVersion
\PrintedOrElectronic

\ifpdf \usepackage[pdftex]{graphicx} \pdfcompresslevel=9
\else \usepackage[dvips]{graphicx} \fi

\usepackage{egweblnk} 

\title{SimJEB: Simulated Jet Engine Bracket Dataset}

\author[E. Whalen A. Beyene C. Mueller]
{\parbox{\textwidth}{\centering E. Whalen, 
A. Beyene,
        and C. Mueller
        }
        \\
{\parbox{\textwidth}{\centering Massachusetts Institute of Technology\\
       }
}
}

\begin{document}

\teaser{
 \includegraphics[width=\linewidth]{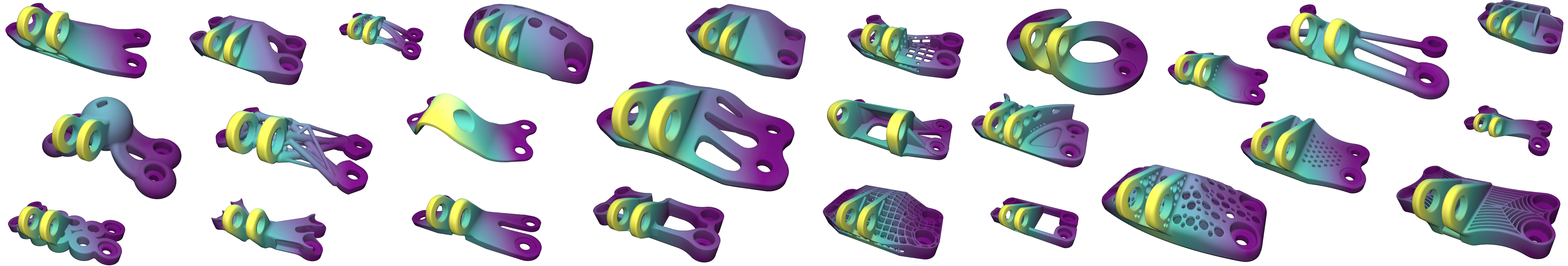}
 \centering
  \caption{Introducing SimJEB: a diverse collection of hand-designed engineering CAD models and accompanying structural simulations (\protect\httpsAddr{//simjeb.github.io/})}

\label{fig:teaser}
}

\maketitle
\begin{abstract}
This paper introduces the Simulated Jet Engine Bracket Dataset (SimJEB) \cite{whalen_simjeb_2021}: a new, public collection of crowdsourced mechanical brackets and accompanying structural simulations. SimJEB is applicable to a wide range of geometry processing tasks; the complexity of the shapes in SimJEB offer a challenge to automated geometry cleaning and meshing, while categorical labels and structural simulations facilitate classification and regression (i.e. engineering surrogate modeling). In contrast to existing shape collections, SimJEB's models are all designed for the same engineering function and thus have consistent structural loads and support conditions. On the other hand, SimJEB models are more complex, diverse, and realistic than the synthetically generated datasets commonly used in parametric surrogate model evaluation. The designs in SimJEB were derived from submissions to the GrabCAD Jet Engine Bracket Challenge: an open engineering design competition with over 700 hand-designed CAD entries from 320 designers representing 56 countries. Each model has been cleaned, categorized, meshed, and simulated with finite element analysis according to the original competition specifications. The result is a collection of 381 diverse, high-quality and application-focused designs for advancing geometric deep learning, engineering surrogate modeling, automated cleaning and related geometry processing tasks. 


\end{abstract}  

\section{Introduction}
Physical simulation plays an important role in designing high-performance engineering components, though the number of required simulations can be computationally prohibitive. Surrogate models, also known as metamodels or response surfaces, are data-driven approximations of computationally-intensive physical simulations that critically accelerate the design process \cite{wang_review_2007, forrester_engineering_2008, queipo_surrogate-based_2005}. While traditional surrogates relied on simple regression models that operated over handcrafted shape parameters, modern surrogates are beginning to leverage geometric deep learning methods for learning directly on 3D shapes \cite{pfaff_learning_2020,baque_geodesic_2018,danhaive_structural_2020, cunningham_investigation_2019}. The implications for engineering design are profound. No longer bound by a parametric design space, this new class of surrogate models has the ability to learn on arbitrary collections of parts. Unfortunately, a lack of quality datasets means that most of these surrogate models are still being evaluated on synthetically generated data and thus not realising their full potential. Furthermore, there does not yet exist a standard benchmark for surrogate modeling for structural engineering applications with which different models can be compared.

Existing shape collections tend to embody one of two extremes. On one hand are synthetically generated datasets, in which a domain expert handcrafts shape procedures or parameters and then randomly samples until the desired number of shapes has been created. While synthetic generation allows for precise control over shape variation, it is challenging to design parameters that will produce both diverse and realistic shapes. As a result, most synthetically generated collections suffer from either excessive homogeneity or a lack of realism. On the other hand are shape datasets collected from various public repositories (i.e. "the wild"). Collected shapes do not suffer from the realism problem and have been invaluable for developing tasks like classification and segmentation; however, the lack of control over shape variation and function typically makes these datasets ill-suited for surrogate modeling, where each shape should be designed for the same engineering task.

This work explores a third source of shape data: online design competitions. Design competition entries occupy the sweet spot between generated and collected datasets. The designs are complex, diverse, and realistic since each one is hand-designed by a different domain expert, and yet each conforms to the functional engineering requirements enforced by the competition. Furthermore, the participating engineers typically design CAD with structural simulation in mind, resulting in cleaner, higher quality CAD models than one might encounter in the wild.

\begin{figure}[htb]
  \centering
  \includegraphics[width=.8\linewidth]{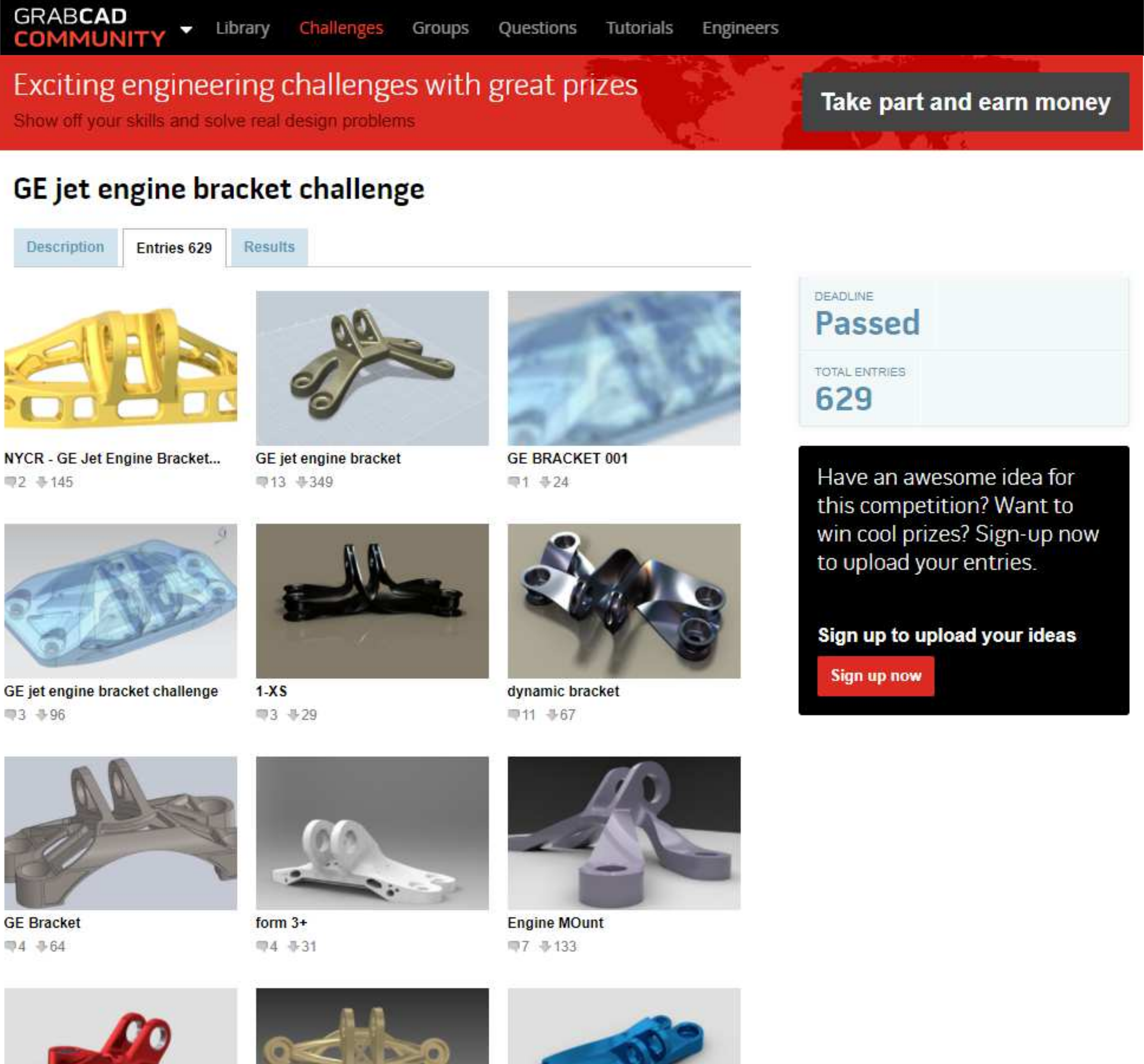}
  \caption{\label{fig:compWebsite}The GE Jet Engine Bracket Competition hosted by GrabCAD.com drew contributions from engineers of many backgrounds and experience levels to compete for cash prizes}
\end{figure}

This paper introduces the Simulated Jet Engine Bracket Dataset (SimJEB) \cite{whalen_simjeb_2021}, a new public shape collection for testing geometric machine learning methods with an emphasis on surrogate modeling (Figure \ref{fig:teaser}). The bracket designs in SimJEB originate from the GE Jet Engine Bracket Challenge: an open engineering design competition hosted in 2013 by GrabCAD.com \cite{kiis_ge_2013} (Figure \ref{fig:compWebsite}). The original competition featured over 700 entries, representing 320 designers from 56 countries whose work is estimated to have taken 14 person-years \cite{morgan_ge_2014}. The diversity of the entries reflects that of their creators, employing a broad range of design strategies, styles and structural behaviors. As mandated by the competition, each bracket has the same four bolt holes and interface point so that they might all be used for the same engineering task. In SimJEB, each design has been cleaned, meshed, and simulated according to the competition's original structural load cases by this papers' authors. For additional analytical potential, each design is labeled as belonging to one of six design categories as determined by a domain expert. Although these particular brackets were designed for use in a jet engine, their structural behavior and design objectives are representative of most structural engineering tasks in civil, mechanical and biomedical engineering. The primary contributions are summarized as follows:
\begin{enumerate}
  \item Introducing SimJEB: a public collection of 381 hand-designed structural engineering CAD models, accompanied by structural simulation results and design category labels, designed for evaluating engineering surrogate models and other geometry processing tasks
  \item Characterizing the collection in terms of geometry, mesh quality, and structural behavior
  \item Proposing a benchmark, including suggested train/test splits, quality metrics, and results from a naive approach, to advance engineering surrogate models and geometric deep learning models
\end{enumerate}

Section \ref{relatedWork} describes how SimJEB differs from existing datasets, section \ref{cleaning} describes the geometry processing pipeline used to create the dataset, section \ref{characterization} characterizes the dataset in terms of shape and structural behavior, section \ref{access} addresses licensing, access and attributions, section \ref{benchmark} proposes how SimJEB might be used as a surrogate modeling benchmark, and section \ref{conclusion} summarizes the conclusions and offers suggestions for future work.

\section{Related work} \label{relatedWork}

\subsection{Synthetically generated shape datasets}
Many techniques exist for generating 3D shapes. The graphics community has made extensive use of procedural models for generating content like buildings \cite{muller_procedural_2006}, spaceships \cite{ritchie_controlling_2015} and indoor scenes\cite{qi_human-centric_2018} (see \cite{shaker_procedural_2016} for a review of procedural modeling techniques). While effective for digital content generation, procedural modeling can be difficult to apply to engineering design where manufacturing constraints or package space requirements necessitate more precise control over allowable shapes. Parametric CAD models allow engineers to precisely explore shape parameters (i.e. a design space) \cite{schulz_interactive_2017}. Parametric models can also be constructed via mesh morphing \cite{lee_multiresolution_1999}, a technique used extensively in mechanical engineering for shape optimization \cite{staten_comparison_2012}. Both procedural and parametric generation require the user to manually codify all of the ways in which shapes may vary. Put eloquently by Krispel et al., "Shape design becomes rule design" \cite{krispel_rules_2014}. This constraint limits the diversity and realism of shapes that can be generated.

More recently, deep learning methods have been used to learn shape generation schemes from a collection of training shapes (see \cite{chaudhuri_learning_2020} for a recent review). Deep shape generation has even been applied to engineering design. Umetani used an autoencoder to learn shape parameters from collected vehicle designs \cite{umetani_exploring_2017}. Oh et al. trained a generative adversarial model on samples from a parametric topology optimization model \cite{oh_deep_2019}. While learning is a promising direction for shape generation, models require large training sets for each new application. Datasets like SimJEB can play a critical role in learning more realistic and practical shape generation models for mechanical design. 

\subsection{Collected shape datasets}
Several large shape datasets have been released in recent years, including the Princeton Shape Benchmark \cite{shilane_princeton_2004}, ModelNet \cite{wu_3d_2015}, ShapeNet \cite{chang_shapenet_2015}, Thingi10k \cite{zhou_thingi10k_2016}, ABC \cite{koch_abc_2019} and Fusion360 \cite{willis_fusion_2020}, as well as tools for visualizing mesh quality like HexaLab \cite{bracci_hexalab_2019}. These resources have been impactful on a wide range of geometry processing tasks, including classification, segmentation, surface normal estimation, automated meshing and cleaning, and shape retrieval (to name a few); however, the lack of accompanying engineering simulations prohibits their use for developing surrogate models. Cunningham et al. performed fluid dynamics simulations on some of the ShapeNet aircrafts and watercrafts but the dataset was not made public \cite{cunningham_investigation_2019}. Aside from the effort required to clean geometry, mesh and simulate, the fundamental challenge with using shape collections in engineering simulations is that the operating conditions for which the part was designed are unknown, and thus it is not possible to characterize its design performance. Schneider et al. used the method of manufactured solutions to solve the Poisson equation on Thingi10k models \cite{schneider_large_2019}. While suitable for PDE discretization studies, the method of manufactured solutions imposes an arbitrary analytical solution and thus does not attempt to capture the physical response of the part under normal operating conditions. In contrast, the specific load and support conditions are known and consistent across all SimJEB models, enabling accurate engineering simulation.

\subsection{Design competition data}
Engineering design competitions have long been a source of innovation (e.g. The Longitude Act of 1714 \cite{burton_prizes_2017} and the Tower of London competition of 1890 \cite{vinnitskaya_tower_2019}). Though modern design competitions, like those hosted by NASA \cite{dunbar_nasa_2019} and ASME \cite{noauthor_asme_nodate}, continue to yield innovative designs, the competitions themselves are a relatively untapped source of functional shape data. Previous works have utilized data from the GE Jet Engine Bracket Challenge, including several that have used the geometry and loads for testing topology optimization methods \cite{gebisa_case_2017, dallash_optimal_2017, muhamad_mass_2020, moosavi_topology_2021}. These works used the competition package space and loads but did not consider the dataset as a whole. Morgan et al. studied 10 of the challenge entries in detail as part of a case study in sustainable design but did not perform physical simulation or release the data \cite{morgan_ge_2014}. McComb et al. trained a voxel-based surrogate model to predict support material and print time for additive manufacturing using 300 voxelized bracket designs \cite{mccomb_predicting_2018}. The data were not made public.

SimJEB is a public collection of 381 cleaned, meshed and simulated designs from the GE Jet Engine Bracket Challenge. The collection is a step towards advancing geometry processing methods for realistic, functional engineering components.

\section{Geometry cleaning and simulation pipeline} \label{cleaning}
\begin{figure*}[htb]
  \centering
  \includegraphics[width=.8\linewidth]{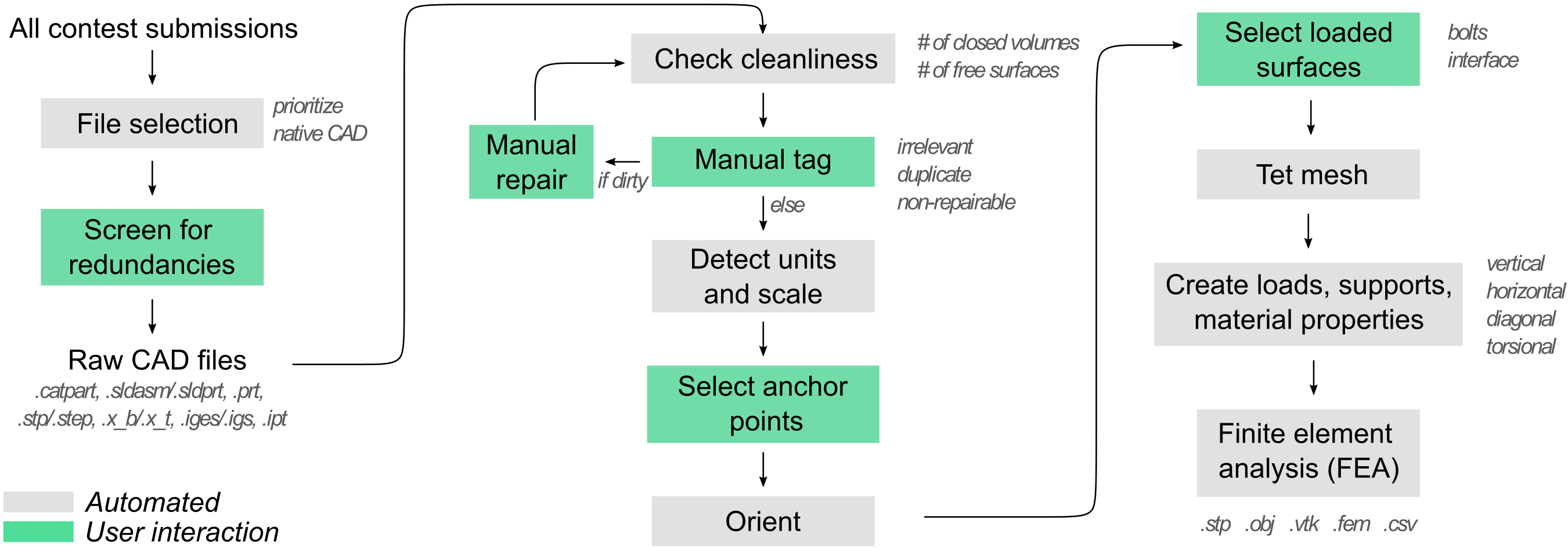}
  \caption{\label{fig:workflow} The semi-automated pipeline used to filter, clean, mesh and simulate the raw CAD contest submissions. Tasks such as orienting, meshing, and checking cleanliness are relatively easy to automate, while tasks that require engineering intuition like assessing part relevancy are best left up to a domain expert.}
\end{figure*}

\begin{figure*}[htb]
  \centering
  \includegraphics[width=.8\linewidth]{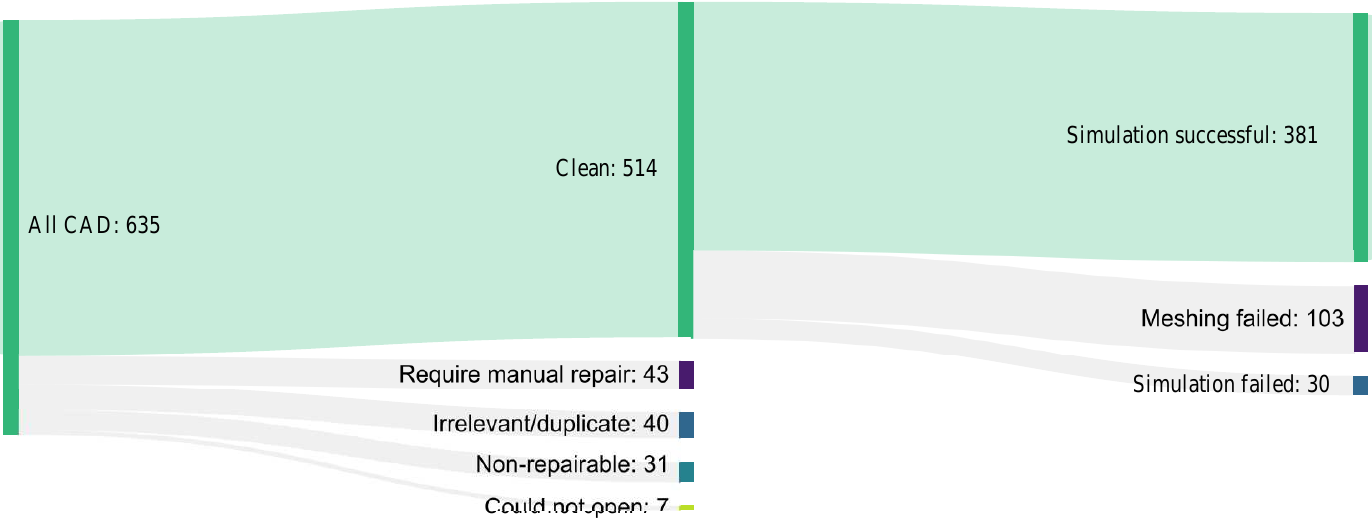}
  \caption{\label{fig:sankey} The "leaks" in the pipeline used to filter, clean, orient, mesh, and simulate bracket models. The clean models include 67 models which were cleaned manually by stitching or deleting surfaces with commercial CAD software. 43 of the models were deemed to be too time consuming to clean by hand.}
\end{figure*}

The following section describes the semi-automated workflow used to acquire, preprocess and simulate each bracket model in SimJEB. The complete workflow is depicted in Figures \ref{fig:workflow} and \ref{fig:sankey}.

\subsection{Design competition overview}
The GE Jet Engine Bracket Challenge was a large engineering design competition hosted by General Electric and GrabCAD.com in 2013 \cite{kiis_ge_2013}. The competition attracted 700 submissions from 320 designers representing 56 countries. By one estimate, the total amount of human-hours required to design the brackets was 700 work weeks or 14 human years \cite{morgan_ge_2014}. Participants were challenged to design the lightest possible lifting bracket for a jet engine, subject to the constraint that the maximum stress in the part did not exceed the yield stress of Ti-6Al-4V titanium over four specified load cases. Entries were limited to designs that fit within a provided package space, and were required to have four bolt holes and an interface hole in specific locations. The bracket designs were also required to be manufacturable via additive manufacturing. Cash prizes totalling to \$30,000 USD were distributed among multiple winners selected by a panel of mechanical engineers from GE and GrabCAD.com. 

\begin{table}
\renewcommand{\arraystretch}{1}
\caption{\label{meshTable}The parameter values use in tetrahedral mesh generation. See \cite{noauthor_altair_nodate-1} for a description of the *tetmesh command and its input parameters.}
\begin{center}
\label{resTable}
\begin{tabular}{l l}
\hline
\textbf{Mesh param} & \textbf{Value} \\ 
\hline
Element type & 1st order trias \\
Element size & 2.0 mm \\
Min size & 0.6 mm \\
Max angle & 30$^{\circ}$ \\
Tet collapse & 0.1 \\
Growth rate & 2.0 \\
QT ratio & 0.8 \\
\hline
\end{tabular}
\end{center}
\end{table}

\subsection{CAD file acquisition}
On the date of access (June 4th, 2020), the GE Jet Engine Bracket Competition website had 629 entries. While most entries contained a single CAD file, some entries contained redundant designs in different CAD file formats, some contained multiple design variations, and still others were missing a CAD file entirely. In the first pass, files were filtered programmatically. If entries contained multiple CAD files with different names (e.g. \emph{"bracket\_v1.stp"}, \emph{"bracket\_v2.stp"}) then both files were retained as they were assumed to be different design variants. If entries contained multiple CAD files with names differing only in the extension (e.g. \emph{"model1.stp"}, \emph{"model1.igs"}), the shapes were assumed to be redundant and only the file with the highest priority extension was retained, where the priority was defined as follows: \emph{.catpart}, \emph{.sldasm}, \emph{.sldprt}, \emph{.prt}, \emph{.stp}, \emph{.step}, \emph{.x\_b}, \emph{.x\_t}, \emph{.iges}, \emph{.igs}, \emph{.ipt}. Note that native formats were preferred over neutral ones. In the second pass, the remaining entries with multiple CAD files were manually screened for obvious redundancies (e.g. \emph{"GE\_Bracket.stp"}, \emph{"GE\_Bracket\_color\_changed.stp"}). At the end of the file selection process, 56 entries had more than one valid CAD file, 518 entries had exactly one, and 55 entries did not have any, resulting in a total of 635 raw CAD files.

\subsection{Geometry cleaning}
Prior to performing structural analysis, each CAD model was cleaned, tagged, scaled, and oriented to a canonical pose through the following semi-automated process. First, an automated check was performed to get the total number of closed volumes in the model. Next, the user was prompted to review the model and optionally assign one of the following tags: \emph{duplicate}, \emph{irrelevant}, \emph{non-repairable}. The model was considered to be clean if it contained exactly one closed volume and was not assigned a tag. 447 of the 635 models were determined to be clean after the first pass. The units of length for each clean model were automatically inferred from the volume of its bounding box and the appropriate scale was applied such that all models were defined in millimeters. The user was then prompted to select three reference points on the model which were used to translate and rotate it to a canonical pose. The models that were considered unclean were either manually cleaned, by deleting extraneous geometry and sliver surfaces or patching non-watertight volumes, or determined to be non-repairable in a second pass, resulting in a total of 514 clean CAD models.

\begin{figure*}[htb]
  \centering
  \includegraphics[width=.8\linewidth]{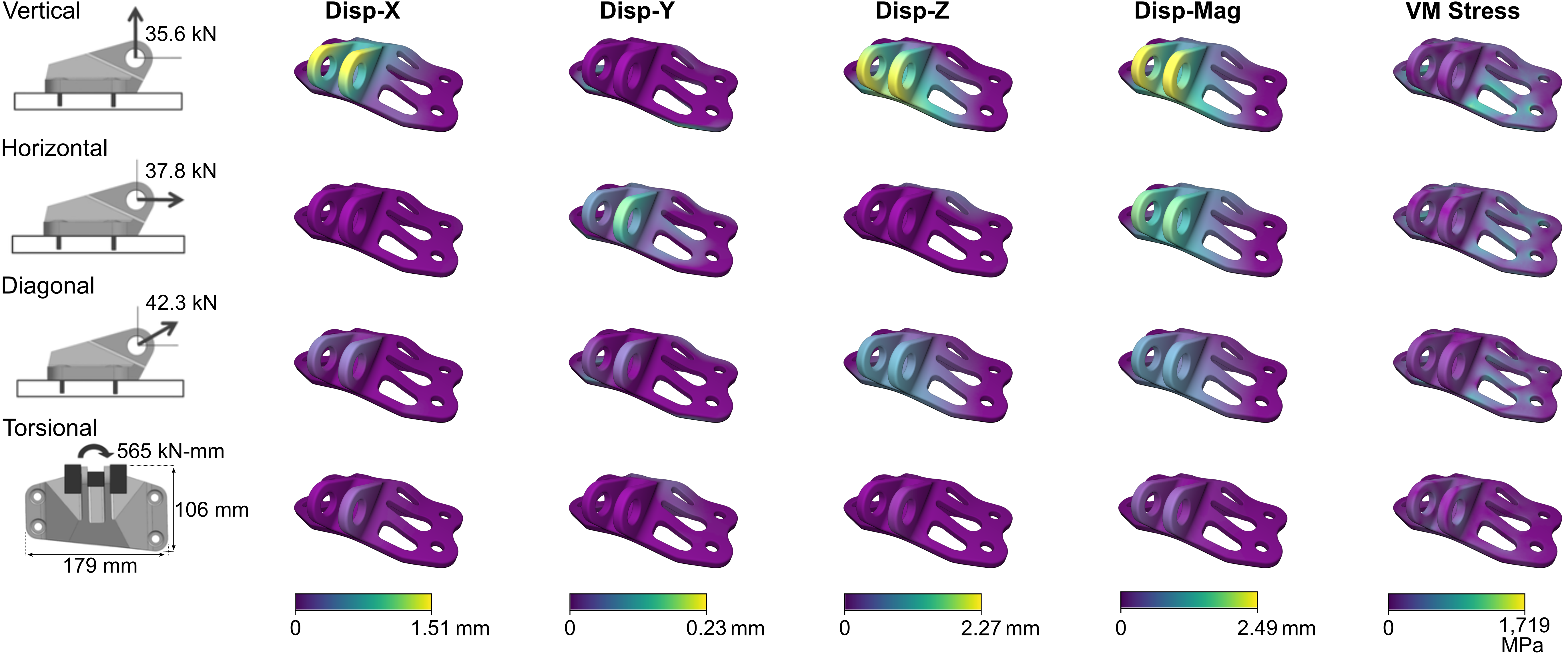}
  \caption{\label{fig:fields} Each bracket was simulated according to the four load conditions specified by the competition. Five vertex-valued scalar fields were extracted for each load case: the displacement in X,Y,Z directions, the displacement magnitude, and the von Mises stress.}
\end{figure*}

\subsection{Finite element structural simulation}
A similar semi-automated process was used to build the finite element models using commercial software \cite{noauthor_altair_nodate-1,noauthor_altair_nodate}. First, the user was prompted to select the surfaces defining all four bolt holes and the interface hole. Next, a first-order tetrahedral mesh was generated with an average element size of 2 mm. The complete list of input parameters used in HyperMesh's \emph{*tetmesh} command can be seen in Table \ref{meshTable}. 411 models were meshed successfully. Meshing was considered to have failed if the resulting mesh violated the 0.1 tet collapse threshold or if the application crashed during meshing. Each bolt was modeled by a rigid RBE2 spider element connected to each mesh node on the selected bolt surfaces and constrained by a Single Point Constraint (SPC) at the center. An RBE3 spider element was used to distribute each of the four loads across the surfaces of the interface hole. As specified in the original competition, the bracket material was modeled as Ti-6Al-4V titanium (\(E=\)113.8 GPa, \(\nu=\) 0.342, \(\rho=\)4.47e-3 g/mm\(^3\)). Finally, each of the four load conditions were simulated using linear-static FEA and the resulting displacements and von Mises stresses were recorded for each node (Figure \ref{fig:fields}). 381 models were simulated successfully. The most common cause of simulation failure was a violation of OptiStruct's mesh quality check, which was left at its default values \cite{noauthor_altair_nodate}. Note that the structural analysis performed for SimJEB may use slightly different assumptions than those of the original competition are are not meant to replace or correct any simulations performed by the original designers. 

\section{Dataset characterization} \label{characterization}
Despite being designed for the same functional purpose, the SimJEB bracket models vary significantly in shape and structural performance. This section characterizes variation in geometry and structural performance across the dataset.

\subsection{Characterization of geometry}
\begin{figure}[htb]
  \centering
  \includegraphics[width=.99\linewidth]{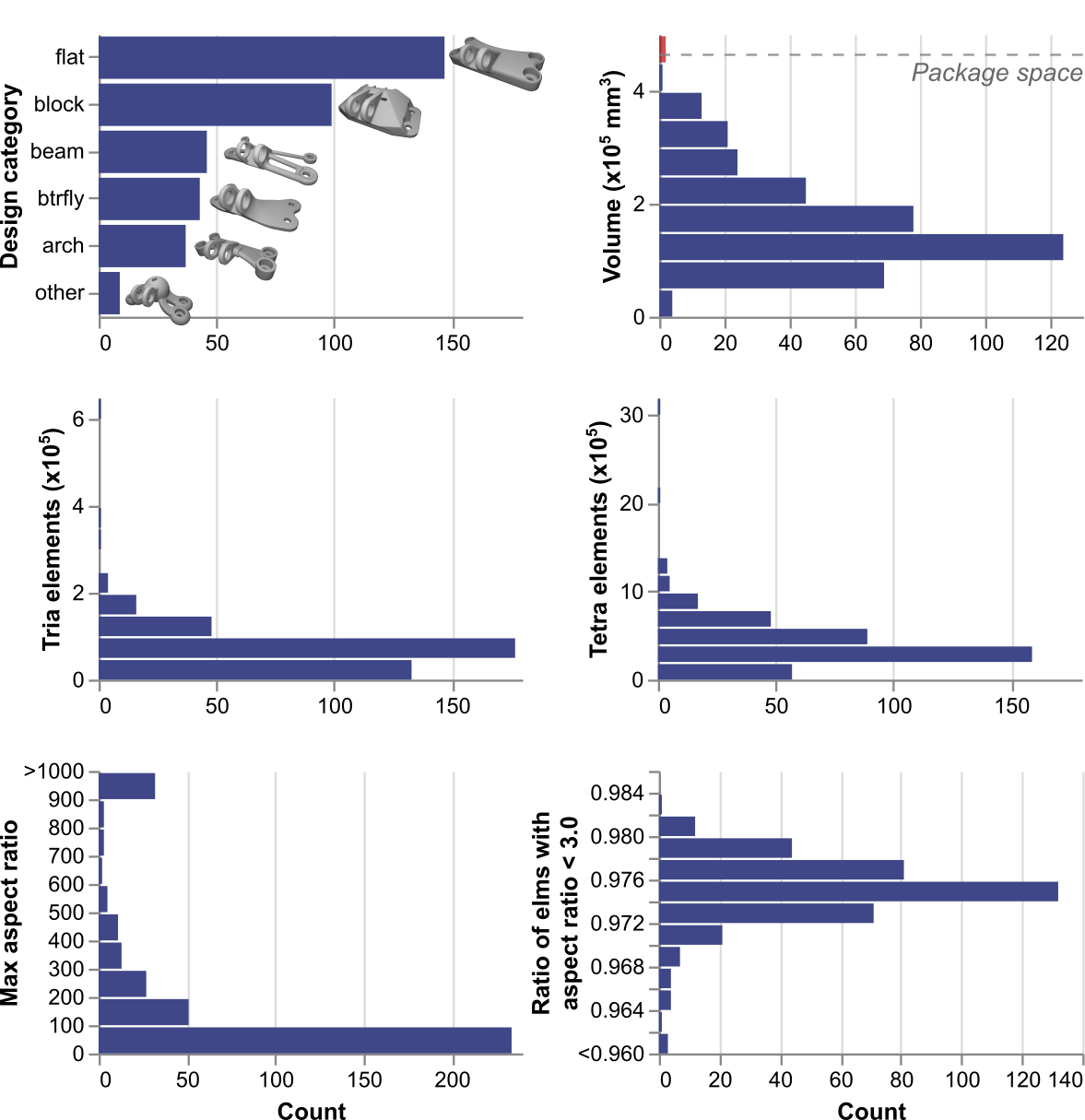}
  \caption{Top left: all brackets are manually labeled as belonging to one of six general design categories. Top right: The volume was bounded above by the competition-specified package space (apart for three designs which violate the rule). Middle: brackets range in the number of triangular and tetrahedral elements in the surface and simulation meshes, respectively. Bottom: tetrahedral mesh quality across the collection measured by aspect ratio.}
  \label{fig:histograms}
\end{figure}

The broad range of designer backgrounds, experience levels, and software tools behind the bracket designs are reflected in the design diversity. While each design has the same bolt holes and interface point mandated by the competition, the remainder of the shape was left to the engineer's imagination. The topology, complexity, and structural design strategy thus vary significantly (Figure \ref{fig:histograms}). In SimJEB, each bracket has been manually assigned to one of six general design categories: \emph{block}, \emph{flat}, \emph{arch}, \emph{butterfly}, \emph{beam} and \emph{other}. \emph{Block} designs were defined as those that occupy a large portion of the allotted package space. \emph{Flat} designs were considered to be those that have mostly flat regions between the bolt holes and interface point, while \emph{arch} and \emph{butterfly} designs have positive and negative curvature in these regions, respectively. \emph{Beam} designs were considered those that have long, slender beam-like regions. Designs that did not fit well into any of these categories were labeled as \emph{other}. The above categorization serves two purposes: 1) it provides a convenient way to partition the data into more homogeneous subsets, and 2) it can be used as labels for testing classification methods. Note that both the definition and assignment of these categories is subjective and imperfect but may be useful in practice for certain modeling or geometry processing tasks.

\begin{figure*}[htb]
  \centering
  \includegraphics[width=.8\linewidth]{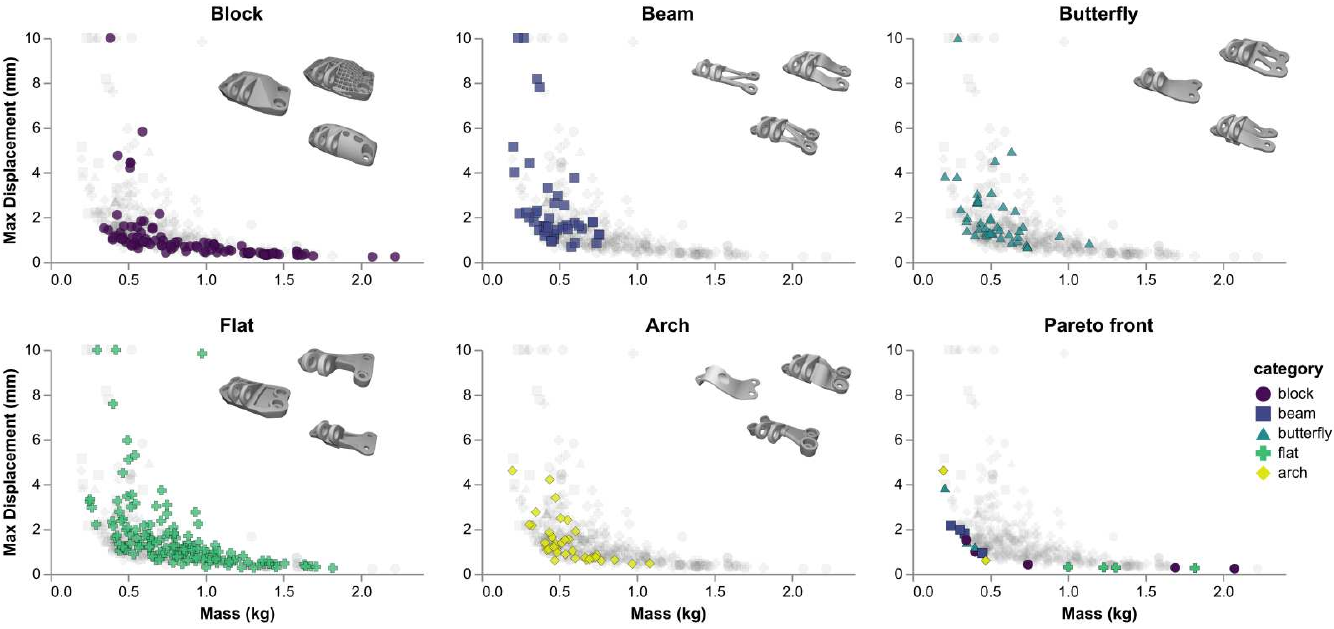}
  \caption{\label{fig:pareto} Multi-objective plots for each of the design categories and for the Pareto optimal designs. Maximum displacements are taken over all mesh vertices and load cases. Optimal designs, which are lightweight and stiff, are located closer to the origin. Note the variety of structural performance within each design category and across the dataset.}
\end{figure*}

\subsection{Characterization of structural performance}
 A common objective in structural engineering is to find lightweight shapes and materials that can withstand specified structural loads. Mechanical failure of metal parts will occur if the maximum von Mises stress at any point inside the part exceeds the known yield stress of the material. Note that the maximum stress does not necessarily lie on the part boundary. A second common objective is to maximize the stiffness of the part, which can be thought of as minimizing the maximum displacement resulting from a given load. Minimizing mass is almost always a competing objective with minimizing displacement and stress, thus the challenge lies in finding (manufacturable) shapes that provide the optimal balance of these quantities for a given application.

The wide variation in SimJEB shapes naturally results in a variety of structural performances. Figure \ref{fig:pareto} shows the performance distribution for each design category in terms of two competition objectives: maximum displacement (over all load cases and vertices) and mass. Designs closer to the origin are lighter and stiffer, and thus more desirable from a structural engineering standpoint. Note that \emph{block} designs tend to be heavier and stiffer, while more minimalist \emph{beam} designs are generally the opposite. \emph{Arch} and \emph{butterfly} designs seem to exhibit the tightest clustering of desirable behaviors. Interestingly, the Pareto front, that is, the set of designs that are optimal for at least one relative weighting of objectives, contains at least one bracket from each design category. 

Simulation accuracy depends in part on mesh quality, which can be challenging to control in an automated pipeline. The majority of SimJEB bracket meshes are comprised of >95\% elements with aspect ratios less than 3.0 (a common industry threshold \cite{lavictor_solidworks_2015}); however, several meshes contain one or more elements with large aspect ratios (Figure \ref{fig:histograms}). While displacement prediction is generally robust to the presence of a few distorted elements, the accuracy of stress predictions could be improved by improving mesh quality, using second-order elements, and replacing sharp corners in the geometry with small fillets.

\begin{figure}[h!]
  \centering
  \includegraphics[width=.8\linewidth]{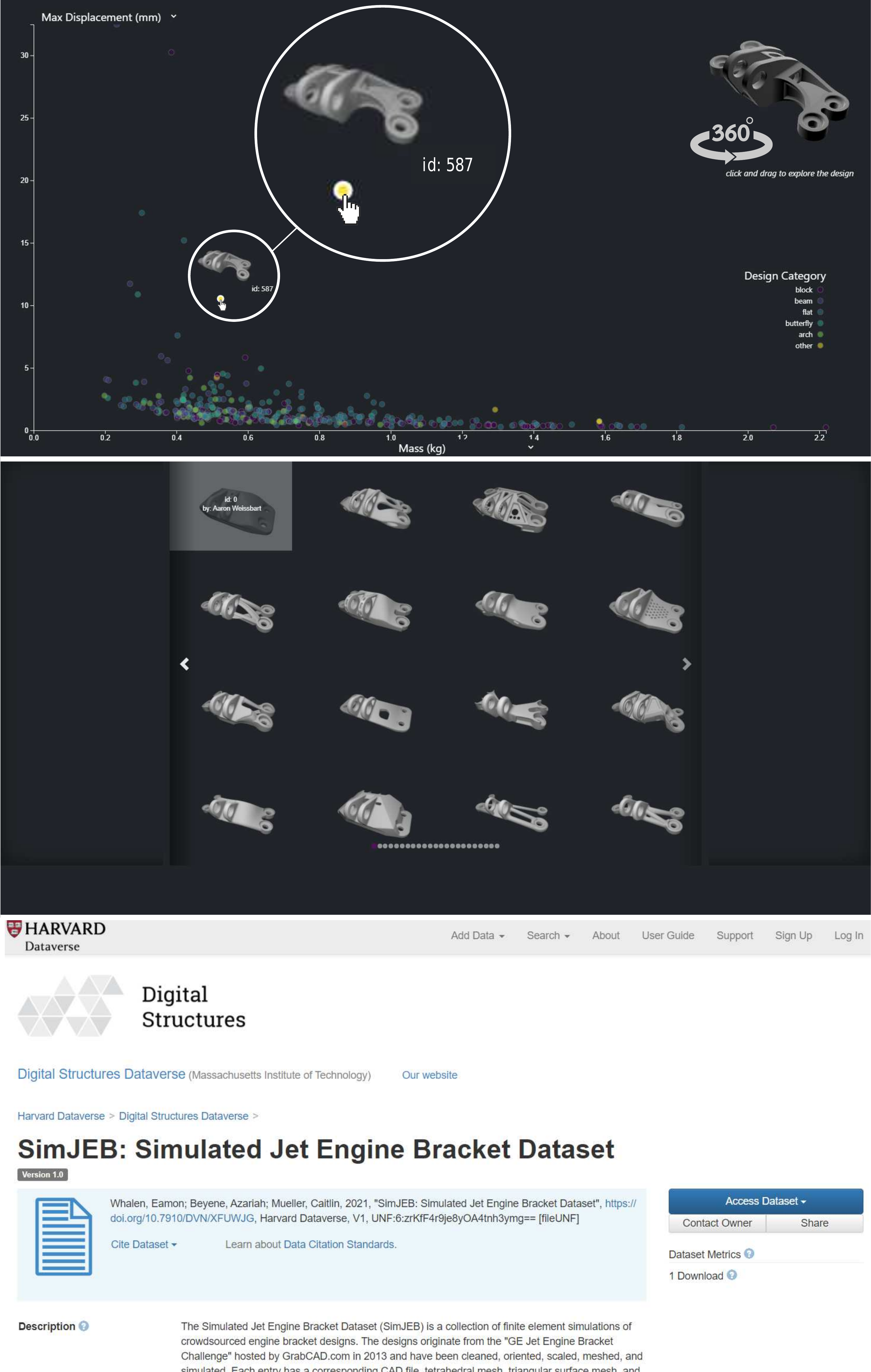}
  \caption{\label{fig:dataverse} Top: An interactive scatter plot of the simulation data with 3D model viewer available at \protect\httpsAddr{//simjeb.github.io}. Middle: A gallery view of all SimJEB models with attributions to their designers. Bottom: The SimJEB dataset is available for public use and hosted through Harvard Dataverse.}
\end{figure}

\section{Licensing, attributions and access} \label{access}
The Simulated Jet Engine Bracket Dataset (SimJEB) is made available under the Open Data Commons Attribution License \cite{noauthor_open_nodate}. CAD files within the database are from the “GE Jet Engine Bracket Challenge”, and licensed for non-commercial use by GrabCAD. All other rights in individual contents of the database are licensed under the Open Data Commons Attribution License.

SimJEB is hosted through Harvard Dataverse and can be accessed at the following link: \protect\httpsAddr{//simjeb.github.io} (Figure \ref{fig:dataverse}). The following data are available for each bracket design: clean CAD (\emph{.stp}), finite element model (\emph{.fem}), tetrahedral mesh (\emph{.vtk}), triangular surface mesh (\emph{.obj}) and simulation results (\emph{.csv}). Each file type is packaged into a separate zip file to facilitate use cases requiring only a subset of file types. A single metadata file provides summary statistics for each bracket and three train/test splits for a benchmark (further explained in section \ref{benchmark}). Additionally, a sample zip file containing one of each file type is provided for convenience. Models are identified by an integer; the files \emph{0.stp}, \emph{0.fem}, \emph{0.vtk}, \emph{0.obj} and \emph{0.csv} thus all belong to model 0. A global README and local README in each zip file attribute each model to its original designer and provide a link to the original submission on GrabCAD.com. All of the original (uncleaned) CAD contest submissions are also available for download with their original submission names to facilitate the testing of geometry cleaning methods or the creation of new versions of SimJEB. Any questions regarding the dataset should be directed to ewhalen@mit.edu.

\section{Surrogate modeling benchmark} \label{benchmark}
Though SimJEB is applicable to a wide range of geometry processing tasks, it was designed primarily for engineering surrogate modeling, that is, learning to predict the displacement (or stress) fields on a 3D shape. The diversity and complexity of the models in SimJEB both exceed those of traditional surrogate modeling datasets, which are almost always synthetically generated. Therefore, SimJEB can be seen as a challenge problem for the learning and structural modeling communities. 

A naive surrogate model is presented to demonstrate how SimJEB might be used as a benchmark. This naive model is simply a degree-three polynomial in the spatial coordinates, with the resulting function approximating the average scalar field across all designs in the training set. Besides demonstrating the benchmark process, this naive model serves as a reference point for predictive performance. Surrogate models that fail to beat this naive model effectively have no predictive value. To standardize training and testing data, three 80/20 train/test splits are provided with the SimJEB metadata. A naive surrogate was trained to predict each of the five scalar fields (\emph{x},\emph{y},\emph{z} components of the displacement, magnitude of displacement, von Mises stress) for each of the four load cases (vertical, horizontal, diagonal, torsional) and for each of the three train/test splits, resulting in 60 surrogate models total. The mean absolute errors (MAEs) in prediction averaged over the three test sets can be seen in Table \ref{benchmarkTable}. Note that MAE is typically preferred over other potential quality metrics because the units are easily interpretable.

\begin{table}
\renewcommand{\arraystretch}{1}
\caption{\label{benchmarkTable}The Mean Absolute Error (MAE) of the naive surrogate model averaged over three train/test splits. These values can be used as a reference point for benchmarking future surrogate models.}
\begin{center}
\begin{tabular}{l l l l l}
\hline
 & \textbf{Vert.} & \textbf{Horiz.} & \textbf{Diag.} & \textbf{Tor.} \\ 
\hline
\textbf{Disp-X (mm)} & 6.27e-2 & 1.62e-1 & 3.17e-2 & 1.27e-1 \\
\textbf{Disp-Y (mm)} & 4.17e-2 & 1.97e-2 & 1.66e-2 & 4.87e-2 \\
\textbf{Disp-Z (mm)} & 1.46e-1 & 1.62e-1 & 2.80e-2 & 2.37e-1 \\
\textbf{Disp-Mag (mm)} & 1.69e-1 & 2.51e-1 & 4.21e-2 & 2.87e-1 \\
\textbf{VM Stress (MPa)} & 6.01e+1 & 8.93e+1 & 3.61e+1 & 8.44e+1 \\
\hline
\end{tabular}
\end{center}
\end{table}

\section{Conclusion} \label{conclusion}
SimJEB is a new collection of realistic, hand-designed engineering models for advancing geometry processing methods. The designs have the same boundary conditions and are accompanied by high-fidelity structural simulation results which makes them ideal for evaluating engineering surrogate models, though the dataset is applicable to a wide range of geometry processing tasks. As the bracket models are hand-designed by structural engineers, they are more realistic and diverse than the synthetically generated datasets. The dataset is characterized in terms of geometry and structural performance, and a benchmark is proposed for surrogate model evaluation. Future work could include improving the robustness of the geometry cleaning and simulation pipeline to increase the total number of bracket designs. Other future projects may include providing RBG-D or multi-view images to support a wider range of learning representations.

\section{Acknowledgements}
Thank you to Bryan Ong and Ashley Hartwell for their excellent work on the SimJEB interactive website. This research was supported by the Engineering Data Science group at Altair Engineering Inc. and is based upon work supported by the National Science Foundation under Grant No. 1854833.

\bibliographystyle{eg-alpha-doi}
\bibliography{references}

\end{document}